\documentclass[aps,prc,twocolumn,superscriptaddress]{revtex4}

\begin{document}


\title{Comment on ``Influence of protons on the capture of electrons by $^7$Be in the Sun''}


\author{B. Davids}
\affiliation{TRIUMF, 4004 Wesbrook Mall, Vancouver, BC, Canada V6T 2A3}
\author{A. V. Gruzinov}
\affiliation{Physics Department, New York University, 4 Washington Place, New York, NY 10003}
\author{B. K. Jennings}
\affiliation{TRIUMF, 4004 Wesbrook Mall, Vancouver, BC, Canada V6T 2A3}


\date{\today}

\begin{abstract}
This paper suffers from conceptual difficulties and unjustified approximations that render its conclusions dubious.
\end{abstract}


\maketitle


A recently published theoretical paper \cite{belyaev07} aims to investigate the influence of protons on the capture of free electrons by $^7$Be in the Sun and other stars. The authors claim that this influence is usually not included in standard treatments of the electron capture, but in fact these effects have been studied comprehensively in the past. We would like to point out that the influence of nearby ions is already included in the standard plasma screening correction to nuclear reaction rates first discovered by Salpeter \cite{salpeter54}, and the same considerations apply to nuclear electron capture in a plasma. The conceptual difficulties and inaccurate approximations contained in this paper vitiate its conclusions.

As explained by DeWitt et al.\ \cite{dewitt73}, the slow pace of nuclear reactions in the solar plasma allows the computation of the screening correction to the reaction rate using equilibrium statistical mechanics. In the case of electron capture, the rate is proportional to the electron density at the position of the $^7$Be nucleus. This density is given by the corresponding average over the Gibbs distribution. Since the surrounding plasma screens the electrostatic potential of the $^7$Be nucleus, the electron-nucleus Coulomb attraction is reduced, decreasing the electron density at the nucleus; the electron capture rate is thus suppressed by the plasma effects.

The primary problem with Ref.\ \cite{belyaev07} is a conceptual difficulty associated with describing the problem. The authors assert that previous investigations have considered only the binary reaction $^7$Be + e$^-$ $\rightarrow ^7$Li + $\nu$, whereas they investigate the ternary reaction $^7$Be + e$^-$ + p $\rightarrow ^7$Li + $\nu$ + p. But this too is a binary reaction, since the proton is a mere spectator that is present in both the initial and final states. Since the nearest proton remains on average some 30 000 fm away from the electron when it is captured by a $^7$Be nucleus at zero range, this proton is well outside of the range of the weak interaction and therefore plays no role in the reaction. In fact, the only influence such a proton can have on the electron capture rate is electromagnetic, by affecting the density of electrons at the $^7$Be nucleus. Therefore it is incorrect to think of this as a ternary reaction. Rather it is a binary reaction that occurs in a plasma environment.

Moreover, the paper treats the case of a three body initial state, $^7$Be + e$^-$ + p. This is an arbitrary and inappropriate choice since there is not merely one neighbouring ion, but several within a distance of 50 000 fm or so,  and of course there are many plasma electrons around as well. These plasma ions and electrons create a fluctuating electric potential at the $^7$Be nucleus which must be evaluated for different possible configurations of the nearby ions and electrons to obtain the average potential. Once calculated, this potential is added to that produced by the $^7$Be nucleus itself and the resulting mean field potential is used to compute the density of electrons at the $^7$Be nucleus. This is a sensible approach to the problem that has been followed by several workers in the field \cite{bahcall62,iks67,bm69,jkl92,bs97,gb97,adelberger98}. It is insufficient to consider a three-body initial state, neglecting the electromagnetic effects of all charged particle spectators except for the nearest proton.

Gruzinov and Bahcall have performed a very careful study of the $^7$Be electron capture rate in the Sun \cite{gb97}. They employed both the usual treatment that divides the electrons into bound and continuum states as well as a density matrix calculation that made no assumptions regarding the quantum states of the electrons in the solar plasma. The calculations included the effects of nearby ions in the plasma, and examined the effects of thermal fluctuations including nonspherical distributions of the neighbouring ions. These density matrix calculations agree very well with the standard mean field approach, within 1\%. The effects of aspherical fluctuations in the ion distribution were also found to be smaller than 1\%. A recent review of $^7$Be electron capture in the Sun \cite{adelberger98} concluded on the basis of these calculations and the others cited above that its rate is known within 2\%.


Electron capture is a two-body reaction. But even if one were to accept the idea that this is a three-body reaction that should be investigated in the context of few-body theory, the approximations of this paper are unjustified. Its equation 2.3 is a poor approximation to the three body wave function 2.2 in the limit of interest, namely when the electron and the $^7$Be nucleus are spatially coincident and the proton is some 30 000 fm away from the other two particles. Clearly, this approximation grows worse and worse as the proton-$^7$Be separation R increases and the magnitude of the Coulomb wave function describing the relative motion of the proton and $^7$Be vanishes. The paper asserts that the Coulomb wave function of the electron in the field of the combined charges of the proton and $^7$Be, $\Psi^{C}(\overrightarrow{r}, Z = Z1 + Z2)$, defines the probability of $^7$Be electron capture. In fact, this is the Coulomb wave function describing the relative motion of an electron and $^8$B, and is only applicable when the proton is closer to the $^7$Be than the electron is. In electron capture this approximation breaks down since the electron-$^7$Be separation must vanish in order for the capture to occur. To set the scale, under solar conditions the mean separation between a $^7$Be nucleus and the nearest proton, some 30 000 fm, is more than 10$^7$ times the Compton wavelength of the W boson, which is roughly the range of the weak interaction that mediates the electron capture. Hence this approximation is invalid.

Moreover, the fact that the wave functions depend on the vectors $\overrightarrow{r}$ and $\overrightarrow{R}$ and not merely their magnitudes is apparently ignored in the paper. There is no discussion of why these Coulomb wave functions do not depend on the relative orientations of the two Jacobi coordinate vectors and not merely on their magnitudes. Also, the fact that the (two-body) Coulomb wave functions depend on the relative energies of the two particles under consideration is ignored. Although the mean thermal energies of three charged particles in a plasma may be the same, this does not imply that the relative energies in the two-body subsystems are identical. Indeed they are not. We found no discussion of these issues in this paper or in Ref.\ \cite{belyaev04}.


In summary, the authors of Ref.\ \cite{belyaev07} assert that three-body processes due to the presence of a proton in the vicinity of the $^7$Be nucleus result in the capture of the electron by an effective charge Z = 5 instead of Z = 4, which is a new effect that cannot be simulated by introducing Debye screening. This is incorrect. For example, the brute-force Monte Carlo simulations of Ref.\ \cite{gb97} compute the electron capture rate without putting in the ion contributions to Debye screening by hand. In these simulations, a proton sometimes appears in the vicinity of the $^7$Be nucleus, yet the resulting plasma screening modification of the electron capture rate is well approximated by the usual Salpeter factor (to within 1\%). There is no contribution from any supposedly new three-body reactions; rather, the electromagnetic effects of plasma electrons and ions of all nuclear species are simultaneously included. Equilibrium statistical mechanics takes care of the three-body and other effects. The standard Salpeter factor provides plasma screening corrections that are sufficiently accurate for solar model calculations \cite{salpeter54,dewitt73,bs97,gb97,gb98}. The fallacious arguments of Ref.\ \cite{belyaev07}, now that they have been answered, are best ignored.

\bibliography{comment}

\end{document}